\documentclass{article}

    \PassOptionsToPackage{numbers, compress}{natbib}

    \usepackage[preprint]{neurips_2024}

\usepackage[utf8]{inputenc} 
\usepackage[T1]{fontenc}    
\usepackage{hyperref}       
\usepackage{url}            
\usepackage{booktabs}       
\usepackage{amsfonts}       
\usepackage{nicefrac}       
\usepackage{microtype}      
\usepackage{xcolor}         
\usepackage{graphicx}
\usepackage{multirow}
\usepackage{caption}
\usepackage{amsmath}
\usepackage{amssymb,amsfonts}        
\usepackage{subfigure}
\usepackage{siunitx}        
\usepackage{makecell}       
\usepackage[ruled,linesnumbered]{algorithm2e}
\usepackage{enumitem}
\usepackage{pifont}     
\usepackage{cleveref}       

\DontPrintSemicolon 

\title{SLIM: Style-Linguistics Mismatch Model for Generalized Audio Deepfake Detection}

\author{%
  Yi Zhu \quad Surya Koppisetti \quad Trang Tran \quad Gaurav Bharaj \\
  Reality Defender \\
  \texttt{\{yi,surya,trang,gaurav\}@realitydefender.ai}\\
}

\begin{document}

\maketitle

\begin{abstract} \label{abstract}
Audio deepfake detection (ADD) is crucial to combat the misuse of speech synthesized from generative AI models. Existing ADD models suffer from generalization issues, with a large performance discrepancy between in-domain and out-of-domain data. Moreover, the black-box nature of existing models limits their use in real-world scenarios, where explanations are required for model decisions. To alleviate these issues, we introduce a new ADD model that explicitly uses the \textbf{S}tyle-\textbf{LI}nguistics \textbf{M}ismatch (SLIM) in fake speech to separate them from real speech. 
SLIM first employs self-supervised pretraining on only real samples to learn the style-linguistics dependency in the real class. The learned features are then used in complement with standard pretrained acoustic features (e.g., Wav2vec) to learn a classifier on the real and fake classes.  
When the feature encoders are frozen, SLIM outperforms benchmark methods on out-of-domain datasets while achieving competitive results on in-domain data. The features learned by SLIM allow us to quantify the (mis)match between style and linguistic content in a sample, hence facilitating an explanation of the model decision.
\end{abstract}    
\section{Introduction} \label{intro}
The growing interest in generative models has led to an expansion of publicly available tools that can closely mimic the voice of a real person~\cite{tan2021survey}.
Text-to-speech (TTS) or voice conversion (VC) systems can now be used to synthesize a fake voice from only a few seconds of real speech recordings~\cite{tan2023neural}.
When these generation tools are used by bad actors, their outputs, commonly referred to as `audio deepfakes'~\cite{khanjani2023audio}, can pose serious dangers. Examples include impersonation of celebrities/family members for robocalls~\cite{news1}, illegal access to voice-guarded bank accounts~\cite{news2}, or forgery of evidence in court~\cite{news3}. Reliable audio deepfake detection (ADD) tools are therefore urgently needed.

State-of-the-art (SOTA) detection systems~\cite{wang2021investigating, yi2023audio} employ self-supervised learning (SSL) encoders as the frontend feature extractors, and append classification backends to map the high-dimensional feature representations to a binary real/fake decision~\cite{li2024cross, yi2023audio, wang2021investigating}.
 Common SSL encoders for this task are the Wav2vec~\cite{baevski2020wav2vec}, WavLM~\cite{chen2022wavlm}, and HuBert~\cite{hsu2021hubert}, among others. 
These models are usually trained in a fully-supervised manner, with fake samples generated using off-the-shelf TTS/VC tools 
~\cite{li2024audio,yi2023audio,chintha2020recurrent, wang2021investigating, muller2023complex, yang2024robust, khan2024frame}. 
However, current ADD systems are known to underperform on deepfakes crafted by unseen generative models (i.e., \textit{unseen attacks})~\cite{liu2023asvspoof, muller2022does, shih2024does, yi2023audio}. To tackle this issue, some works have focused on extracting more robust features from the input representation~\cite{jung2022aasist,tak2022automatic,yang2024robust}. Additional improvements have been reported by finetuning the SSL frontend during downstream supervised training~\cite{tak2022automatic, wang2021investigating, 9747768} and by increasing the diversity of labelled samples via data augmentation or continual training on vocoded data~\cite{tak2022rawboost, xie2023learning, wang2023spoofed, wang2024can}. While shown to be effective for in-domain deepfakes, frontend finetuning increases the cost of training drastically.

Additionally, outputs from existing ADD systems are hard to explain, i.e., it is unclear to a typical user why an ADD makes a certain prediction, which leads to lack of trust~\cite{yi2023audio, li2024audio}. For practical applications, it is crucial to understand what information the model is relying on to make decisions, and under which circumstances the model would fail to successfully detect deepfakes. A group of works uses explainable AI (xAI) methods~\cite{arrieta2020explainable} to interpret model decisions~\cite{ge2022explaining, kwak2023voice, lim2022detecting}, but they mainly rely on {\em post-hoc} visualizations such as saliency maps, which are known to be sensitive to training set-ups~\cite{yanagawa2023seeing} and can therefore be inconsistent. Other models focus on specific vocal attributes, such as breath~\cite{layton2024every}, or vocal tract~\cite{blue2022you} to derive explanations. However, most of these interpretable attributes only account for a subset of deepfake-related characteristics, hence resulting in a large gap in detection performance compared to SOTA methods~\cite{yi2023audio,li2024audio}. We note that while ``interpretability/explainability'' is often ambiguous \cite{lipton2018}, in this work we mean the model's ability to provide reasons for a certain prediction, e.g., a sample is likely fake because its style and linguistics representations are more misaligned than those of real samples (as will be shown in Section \ref{explain}).

In this study, we propose a generalizable ADD model that explicitly explores the style-linguistics mismatch in deepfakes to separate them from real ones, and thereby facilitates an explanation on the model decision. We hypothesize that in real speech, a certain dependency exists between the \emph{linguistics} information embedded in the verbal content and the \emph{style} information embedded in the vocal attributes, such as speaker identity and emotion. To synthesize a deepfake audio, both TTS and VC systems artificially combine the verbal content with the vocal attributes of a target speaker, and thereby introduce an artificial style-linguistics dependency that would differ from the real speech. Our two-stage framework explicitly studies the Style-LInguistics Mismatch (SLIM) in the fake class to separate it from the real class. During Stage 1, the style-linguistics dependency in the real class is learned by contrasting the style and linguistic subspace representations and generating a set of dependency features from each subspace. The learned pairs of style and linguistics features are expected to be more correlated for real speech than for deepfake speech. In Stage 2, we employ supervised training, wherein we fuse the learned dependency features from Stage 1 with the original style and linguistic representations and train a light-weight projection head to classify the input representations as real or fake.

Our main contributions are summarized as follows:
\begin{enumerate}[partopsep=0pt,topsep=0pt,parsep=0pt,leftmargin=*]
    \item We propose SLIM, a model leveraging Style and LInguistics Mismatch in deepfake audios for generalized detection with interpretation capabilities.  
    \item SLIM outperforms existing SOTA methods on out-of-domain datasets (\texttt{In-the-wild}, \texttt{MLAAD}) while being competitive on in-domain datasets (\texttt{ASVspoof2019},  \texttt{2021}). This is achieved without increasing the amount of labeled data or the added cost from end-to-end finetuning.
    \item Unlike black-box ADD models, the style-linguistics features learned by SLIM can be used to interpret model decisions. We present analyses to show how the interpretation can be performed on a group level as well as on individual speech samples.

\end{enumerate}

\section{Related works}
\label{sec:relatedworks}
\subsection{Audio deepfake detection}
State-of-the-art ADD systems mainly rely on fully-supervised training, where the model architectures comprise of one or more speech SSL frontends and a backend classifier~\cite{yi2023audio, almutairi2022review, masood2023deepfakes}. \citet{guo2024audio} developed a multi-fusion attentive classifier to process the output from a WavLM frontend; \citet{yang2024robust} fused outputs from multiple SSL frontends and reported improvements over using a single frontend. However, existing ADD systems experience severe degradation in performance when tested on unseen data~\cite{muller2022does, shih2024does}, which questions their applicability and trustworthiness for real-world scenarios. To address this issue, multiple works have explored methods to improve model generalizability. With added training cost, improvements have been reported when frontends are finetuned alongside the backend classifiers during downstream training~\cite{tak2022automatic, wang2021investigating}. Further improvements were achieved with data augmentations such as RawBoost~\cite{tak2022rawboost, tak2022automatic} and neural vocoding~\cite{wang2023spoofed}. More recent works also show that distilled student models can generalize better than large teacher models~\cite{lu2024one, wang2024can}. Still, large discrepancies between in-domain and out-of-domain performance are common~\cite{yi2023audio,li2024audio}.

In addition to generalization, existing ADD models also fall short on interpretability. Several studies have shown that current SOTA models may be focusing on artifacts introduced in the frequency spectrum during voice synthesis  and/or the artifacts in non-speech segments~\cite{shih2024does, muller2021speech, liu2023asvspoof, zhang2023impact}. 
While a line of work proposed to extract speech-related features, such as breath~\cite{layton2024every} and vocal tract and articulatory movement~\cite{blue2022you}, the overall detection performance was inferior to SSL-based methods. Other works resort to xAI methods~\cite{arrieta2020explainable} for model interpretation, such as SHAP~\cite{ge2022explaining}, GradCAM~\cite{kwak2023voice}, and Deep Taylor~\cite{lim2022detecting}. However, these post-hoc analysis approaches are known to be sensitive to training set-ups~\cite{yanagawa2023seeing} and therefore not viable for practical use. Both generalization and interpretability remain challenging issues for current ADD systems.

\subsection{Style-linguistics modelling}
One standard approach for modelling speech is to decompose it into two subspaces, style and linguistics. The former refers to short and long-term paralinguistic attributes, such as speaker identity, emotion, accent, and health state~\cite{schuller2013paralinguistics}. The latter corresponds to the verbal content of speech~\cite{kretzschmar2009linguistics}. For representing style information, early works relied on handcrafted features, such as GeMAPS~\cite{eyben2015geneva,eyben2010opensmile}. Later studies showed improved performance by representations learned end-to-end by deep neural networks (DNN), such as the x-vector~\cite{snyder2018x} and ECAPA-TDNN embeddings~\cite{desplanques2020ecapa}. Similarly, the linguistic representations follow a similar trend where DNN-based embeddings, such as Whisper~\cite{radford2022whisper}, outperform handcrafted features for content-related tasks~\cite{dhanjal2024comprehensive}. More recent studies have shown that style and linguistics information can be efficiently encoded together in the SSL representations~\cite{baevski2020wav2vec, chen2022wavlm, hsu2021hubert}. To investigate how speech information is encoded in DNNs, a group of works conducted layer-wise analysis and showed that early to middle layers carry more style related attributes, such as speaker identity~\cite{ashihara2024self}, 
emotion~\cite{saliba2024layer}, and articulatory movement~\cite{cho2023evidence}; while later layers encode linguistics attributes, such as phonetic information and semantics~\cite{pasad2023comparative, shah2021all}.

Despite these approaches, it is unclear if completely disentangling style and linguistics information in speech is possible. Studies have shown that a certain dependency exists between these two subspaces: the link between emotional states and word choices~\cite{lindquist2015role}, the relation between prosody and language understanding~\cite{cutler1997prosody}, and the impact of age on sentence coherence~\cite{pereira2019age}. Effectively modeling both the independent and dependent aspects of style and linguistics in speech still remains a challenge.

\section{Method} \label{method}
\subsection{Motivation}
\label{ssec:prelim}
For the majority of generative speech models, the style and linguistic subspaces are assumed to be independent of each other~\cite{tan2021survey, kaur2023conventional, triantafyllopoulos2023overview, mohammadi2017overview}. For example, VC systems change the voice of an utterance by replacing the source speaker's embeddings with those of the target speaker~\cite{mohammadi2017overview, triantafyllopoulos2023overview}, assuming that these embeddings contain no linguistics information. Similarly, modern TTS systems rely on independently learned representations to model different speech aspects (e.g., text, speaker, emotion) to synthesize expressive speech~\cite{baevski2022data2vec, desplanques2020ecapa, triantafyllopoulos2024expressivity}.

Because of this disentanglement assumption, a mismatch likely exists between the style and linguistics information in TTS/VC speech that differentiates it from real speech. To study this hypothesis, we conduct a proof-of-concept experiment on a sample subset of \texttt{ASVSpoof2019}~\cite{todisco2019asvspoof}. 
Following previous research~\cite{raghu2017svcca, kornblith2019similarity, pasad2021layer}, we use canonical correlation analysis (CCA) to derive a subspace where the linear projections of the style and linguistics embeddings are maximally correlated for the real class.
We choose the last layer output of pretrained \texttt{wav2vec2-large-xlsr-53-english}~\cite{grosman2021xlsr53-large-english} for linguistics representation, and the pretrained \texttt{ECAPA-TDNN} embeddings~\cite{desplanques2020ecapa} for style representation. 

\setlength{\tabcolsep}{4pt}
\begin{table}[hbt]
\caption{Mean and standard deviation of Pearson correlation coefficients (\textit{r}) calculated between style and linguistics embeddings for real and TTS/VC samples across 5 unseen speakers. 
Significant difference (calculated by Welch's t-test) is seen between real speech and all types of generated speech.}
\label{tab:cca}
\centering
\begin{tabular}{cccccccc}
\toprule
\bf{Class} & Real & A01 (TTS) & A02 (TTS) & A03 (TTS) & A04 (TTS) & A05 (VC) & A06 (VC)\\
\midrule
\bf{\textit{r}} & {\bf{.308$\pm$.025}} & .202$\pm$.033 & .217$\pm$.020 & .243$\pm$.024 & .253$\pm$.021 & .214$\pm$.026 & .252$\pm$.020 \\ 
\bottomrule
\end{tabular}
\end{table}

We randomly select 100 real speech samples from \texttt{ASVspoof2019} \cite{todisco2019asvspoof} training set to fit 20-dim CCA features for both linguistics and style representations. We then apply the CCA projection to 200 audios from 5 unseen speakers and 6 TTS/VC systems, and compute the correlation values between these projected style and linguistics vectors to quantify the subpace similarities. 

Table.~\ref{tab:cca} shows these results. A higher \textit{r} is seen for the real samples, whereas significantly lower correlations are observed for both TTS and VC generated samples. Moreover, TTS-samples on average show lower \textit{r} (0.228) than VC-samples (0.236), indicating that VC-samples are closer to real speech in terms of style-linguistics dependency. This could explain why VC samples were found to be more challenging to detect than TTS samples in the ASVspoof2019 challenge~\cite{liu2023asvspoof}.
While our findings demonstrate the usefulness of CCA for validating the subspace mismatch, its limitations, such as that it only explores the linear composites of the variables~\cite{weenink2003canonical}, might make it sub-optimal to be used independently for deepfake detection. We therefore develop a detection framework that {\em explicitly} studies the style-linguistic mismatch and scales to larger amount of data.

\subsection{Formulation of SLIM}
\begin{figure}[ht]
\centering
\includegraphics[width=0.8\linewidth]{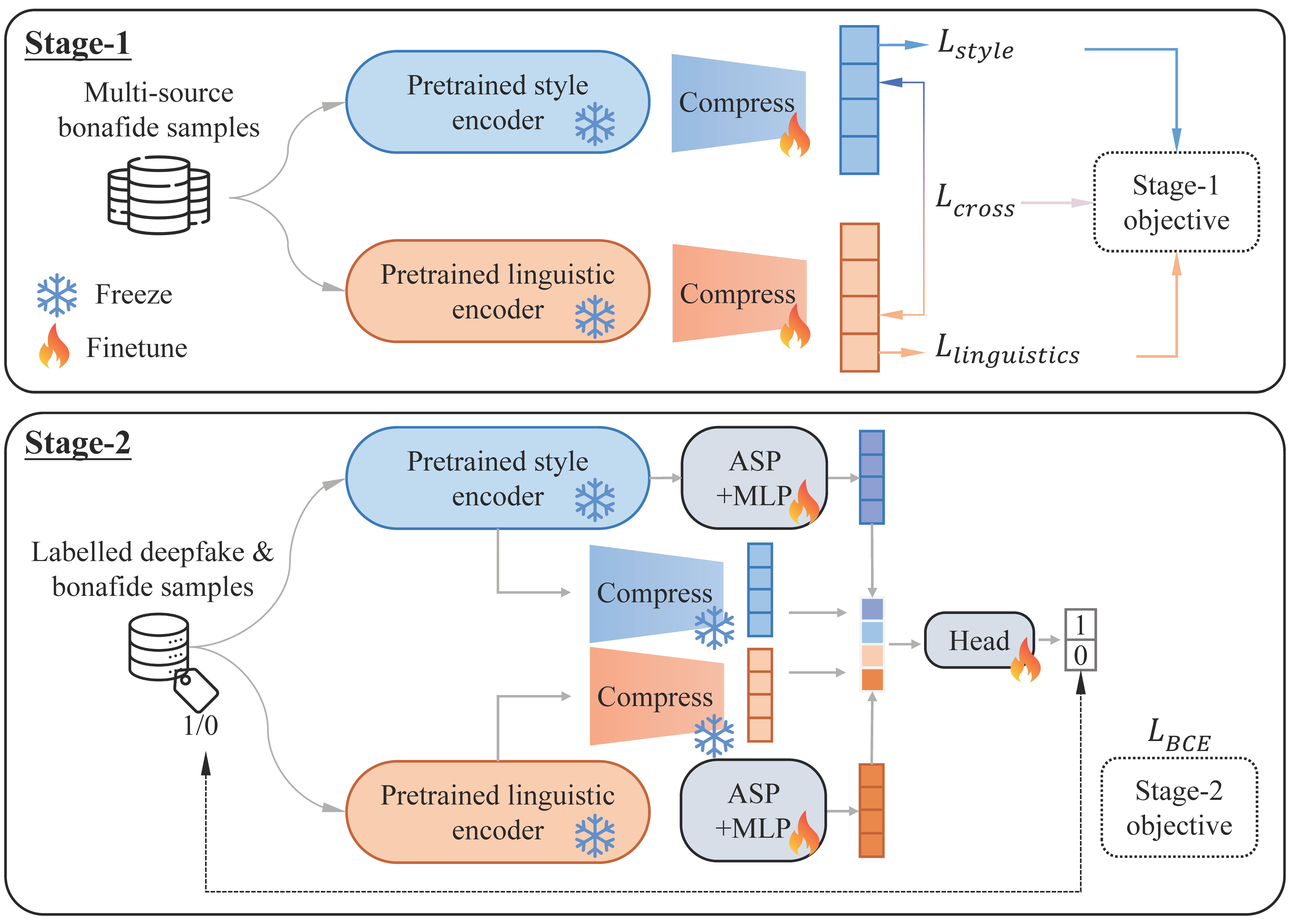}
\caption{SLIM: A two-stage training framework for ADD. Stage 1 extracts style and linguistics representations from frozen SSL encoders, compresses them, and aims to minimize the distance between the compressed representations ($\mathcal{L}_{cross}$), as well as the intra-subspace redundancy ($\mathcal{L}_{style}$ and $\mathcal{L}_{linguistics}$). The Stage 1 features and the original subspace representations (pretrained SSL embeddings) are combined in Stage 2 to learn a classifier via supervised training.
}
\label{fig:training}
\end{figure}

Our two-stage Style-LInguistics Mismatch (SLIM) learning framework is outlined in Fig.~\ref{fig:training}. The first stage operates on the real class only and employs self-supervised learning to build style and linguistic representations and their dependencies for real speech. In the second stage, a classifier is fit onto the learned representations via supervised training over deepfake datasets with binary (real/fake) labels. 

\subsubsection{Stage 1: One-class self-supervised contrastive training} \label{subsec:stage1}
The goal of the first stage is to learn pairs of dependency features from style and linguistics subspaces, which are expected to be highly correlated for real samples and minimally correlated for deepfakes. Since only real samples are needed, we incorporate 
other open-source speech datasets to diversify the style variations. Given a speech sample, we first extract the style and linguistics representations separately using pretrained networks. 
Since recent SSL models achieve superior performance on multiple speech downstream tasks compared to conventional speech representations (e.g., ECAPA-TDNN)~\cite{baevski2020wav2vec, chen2022wavlm, hsu2021hubert, yang2024large,pepino2021emotion} we select a group of SSL models finetuned for paralinguistics and linguistics tasks as candidate encoders~\cite{fan2021exploring, vaessen2022fine, pepino2021emotion, wang2021fine, babu2021xls}. 
In addition, it has been shown that early to middle model layers carry paralinguistics information, while later layers encode linguistics content~\cite{pasad2021layer, ashihara2024self, lin2023utility, shah2021all}; we conducted thorough analyses to examine the cross-correlation between pretrained SSL model layers (Appendix~\ref{appendix:1}) and chose layer 0-10's output from \texttt{Wav2vec-XLSR} fine-tuned for speech emotion recognition to represent style, and layer 14-21's output from \texttt{Wav2vec-XLSR} finetuned for automatic speech recognition, to represent linguistics information.

Both style ($\mathbf{X}_S$) and linguistics ($\mathbf{X}_L$) embeddings are three-dimensional tensors $\in \mathbb{R}^{K\times F \times T}$ 
where $K$ denotes the transformer layer index, $F$ denotes the feature size, and $T$ denotes the number of time steps. 
These subspace embeddings are sent into compression modules $\mathcal{C}(\cdot)$, which average the transformer layer outputs and reduce the feature size from 1024 to 256 (see also Appendix~\ref{appendeix:comp}). We refer to the output from the compression modules as dependency features: $\mathbf{S}_{f,t} = \mathcal{C}(\mathbf{X}_S)$ for style and $\mathbf{L}_{f,t} = \mathcal{C}(\mathbf{X}_L)$ for linguistics, and their temporally averaged versions are denoted $\bar{\mathbf{S}}_{f}$ and $\bar{\mathbf{L}}_{f}$. These dependency features are learned by minimizing the self-contrastive loss $\mathcal{L}_{con}$, defined as: 

\begin{align}
\begin{split}
    \mathcal{L}_{con} = \mathcal{L}_{cross} + \mathcal{\lambda L}_{intra}, & \qquad  \mathcal{L}_{intra} = \mathcal{L}_{style} + \mathcal{L}_{linguistics} 
\end{split} \label{eq:1}\\
\begin{split}
    \mathcal{L}_{cross} = \frac{1}{T} \sum_{t=0}^{T}\|\mathbf{S}_{f,t}-\mathbf{L}_{f,t}\|^{2}_{\mathbf{F}}, & \qquad \mathcal{L}_{intra}  = \|\mathbf{\bar{S}}_{f}\mathbf{\bar{S}}^\intercal_{f}-\mathbb{I}\|^{2}_{\mathbf{F}} + \|\mathbf{\bar{L}}_{f}\mathbf{\bar{L}}^\intercal_{f}-\mathbb{I}\|^{2}_{\mathbf{F}}
\end{split} \label{eq:2} 
\end{align}
\noindent $\mathcal{L}_{cross}$ denotes the cross-subspace loss; $\mathcal{L}_{intra}$ is the intra-subspace loss, defined in terms of $\mathcal{L}_{style}$ and $\mathcal{L}_{linguistics}$ (Figure~\ref{fig:training}); 
$\lambda \in [0,1]$ is a hyperparameter that weighs the two loss terms, $T$ is the number of time steps; and $\|(.)\|^{2}_{\mathbf{F}}$ is the Frobenius norm. The $\mathcal{L}_{cross}$ term reduces distance between the compressed style and linguistic embeddings, while the $\mathcal{L}_{intra}$ term reduces redundancy within the (temporally averaged) style and linguistic features by pushing off-diagonal elements to zero.  

The learned dependency features from Stage 1 can be used to quantify whether a mismatch exists between the style and linguistics of an audio. We further demonstrate this in Section~\ref{mismatch}.

\subsubsection{Stage 2: Supervised training}
The second stage of SLIM follows a standard supervised training scheme, where the dependency features and subspace representations are concatenated and fed into a classification head to generate a binary real/fake outcome. As shown in Figure~\ref{fig:training}, the subspace SSL encoders and compression modules are obtained from Stage 1 and are all frozen during Stage 2. 
Since the dependency features are specifically designed to capture the style-linguistics mismatch alone, we complement them with the original embeddings in order to capture other artifacts that can help separate real samples from the fake class. The original embedding's dimensions are reduced from 1024 to 256 through an attentive statistics pooling (ASP) layer and a multi-layer perceptron (MLP) network. The projected subspace embeddings when concatenated with dependency features result in 1024-dim vectors. The classification head consists of two fully-connected layers and a dropout layer. Binary cross-entropy loss is used to jointly train the ASP and MLP modules alongside the classification head.

\section{Experiments}
Based on the preliminary results from Section \ref{ssec:prelim}, we systematically assess the in-domain and cross-domain detection performance of SLIM using multiple datasets, and demonstrate how such framework would benefit the interpretation of model decisions.

\subsection{Experimental set-up} \label{setup}
\paragraph{{Stage 1 training.}} Unlike benchmark models which are trained end-to-end in a supervised manner, our model relies on two-stage training where each stage requires different training data to avoid information leakage. Since only real samples are needed in Stage 1, we take advantage of open-source speech datasets by aggregating subsets from the \texttt{Common Voice}~\cite{ardila2020common} and \texttt{RAVDESS}~\cite{livingstone2018ryerson} as training data and use a small portion of real samples from the \texttt{ASVspoof2019 LA train} for validation. Both \texttt{Common Voice} and \texttt{RAVDESS} cover a variety of speaker traits. The former is a crowdsourced dataset collected online from numerous speakers with uncontrolled acoustic environments, while the latter is an emotional speech corpus with large variations in prosodic patterns. Such data variety enables our model to learn a wider range of style-linguistics combinations.

\paragraph{{Stage 2 training and evaluation.}}
For a fair comparison with existing works, we adopt the standard train-test partition, where only the \texttt{ASVspoof2019} logical access (LA) training and development sets are used for training and validation. For evaluation, we use the test split from \texttt{ASVspoof2019 LA}~\cite{todisco2019asvspoof} and \texttt{ASVspoof2021 DF}\cite{liu2023asvspoof}. \texttt{ASVspoof2019 LA} and \texttt{ASVspoof2021 DF} have been used as standard datasets for evaluating deepfake detection models, where real speech recordings originate from the \texttt{VCTK} and \texttt{VCC} datasets~\cite{VCTK, lorenzo2018voice, yi2020voice} and the spoofed ones are generated with a variety of TTS and VC systems. Compared to \texttt{ASVspoof2019 LA}, \texttt{ASVspoof2021 DF} contains more than 100 different types of generated speech in the evaluation set, providing a more stringent setting for testing generalization to unseen attacks.
In addition, we assess our model's generalizability on out-of-domain data: \texttt{In-the-wild}~\cite{muller2022does}, and the English subset from \texttt{MLAAD v3}~\cite{muller2024mlaad}.  
\texttt{In-the-wild} consists of on audio clips collected from English-speaking celebrities and politicians, featuring more realistic and spontaneous speech samples. The English subset of \texttt{MLAAD} (hereinafter referred to as \texttt{MLAAD-EN}) is a recent dataset with spoofed samples generated using state-of-the-art open-source TTS and VC systems (more details in Appendix~\ref{appendix:3}). 

\paragraph{{Metrics.}} Equal error rate (EER) is a standard metric for evaluating deepfake detection systems~\cite{todisco2019asvspoof, liu2023asvspoof}. It refers to the point in the detection error tradeoff curve where the false acceptance rate equals the false rejection rate. Lower EER suggests better performance. We also report F1-beta scores ($\beta$=1) to account for the class imbalance. Higher F1 scores suggest better performance.

\paragraph{{Benchmarks.}} \label{benchmark}
We consider several SOTA models to benchmark against and broadly categorize them as follows, based on the training cost: (i) methods which freeze feature extraction frontends and finetune only the backend classifiers, and (ii) methods which finetune frontends together with the classifiers during supervised training. As benchmarks representing the former case, we consider Wav2vec-XLSR+LLGF (W2V-LLGF)~\cite{xie2023learning}, Wav2vec-XLSR+LCNN (W2V-LCNN)~\cite{xie2023learning}, six different models that share a similar backend classifier as SLIM (W2V/WLM/HUB-ASP), a model that fuses different SSL representations (SSL-fusion)~\cite{yang2024robust}, as well as three methods that do not rely on large SSL encoders, namely, LCNN~\cite{chintha2020recurrent}, RawNet2~\cite{tak2021end}, and PS3DT~\cite{yadav2023ps3dt}. For the end-to-end fine-tuning benchmarks, we consider the model in \cite{9747768} with a backend classifier similar to SLIM's (W2V-ASP-ft), and the model in~\cite{tak2022rawboost} with RawBoost augmentation and
AASIST backend (W2V-AASIST).  
Using frozen frontends, five variants of SLIM are considered, where the input at Stage 2 is: (i) only the style embedding, (ii) only the linguistics embedding, (iii) the combination of style and linguistics embeddings, (iv) only the style-linguistics dependency features, and (v) the fusion of style and linguistic embeddings and their dependency features. We emphasize that the original SLIM framework does not involve any finetuning of frontends, since the finetuning may change the disentanglement of style and linguistics embeddings and thus hamper model explainability. However, to compare with finetuned benchmarks, we include a variant of SLIM that finetunes all modules during Stage 2, noting that this would compromise the feature interpretation.

\paragraph{{Implementation details.}} \label{impl}
We implement our models using the SpeechBrain toolkit~\cite{ravanelli2021speechbrain} v1.0.0. The hyperparameters used for Stage 1 and Stage 2 training are provided in Appendix~\ref{appendix:4}. When setting up our customized benchmark models, we followed consistent training recipes where only the model architectures were changed and the same data augmentation method was used. Each round of evaluation was repeated three times with different random seeds, and the mean values are reported. 
\subsection{Experiment results}

\paragraph{Detection performance.} \label{frozen}
Table~\ref{tab:performance_frozen} summarizes the detection performance of all models and compares the number of trainable parameters. We discuss the models with \emph{frozen frontend} here, and compare the models with \emph{finetuned frontend} in Section.~\ref{ablation}. \texttt{ASVspoof2019} \texttt{eval} set contains 19 types of attacks, out of which 6 are seen during training. This makes it the simplest of the four test datasets. We see that a majority of the models achieve near-perfect performance, with several including SLIM reporting EER below 1\%. As expected, degradation is seen when models are tested on \texttt{ASVspoof2021}, where the majority of attacks are unseen. Both W2V-LCNN and SLIM are top-performers, with no significant difference between the two. With the out-of-domain datasets (\texttt{In-the-wild} and \texttt{MLAAD-EN}), more severe degradation is observed, where the majority report EERs over 20\%. SLIM, however, outperforms the others with EER of 12.9\% and 13.5\% on \texttt{In-the-wild} and \texttt{MLAAD-EN}, respectively. It should be noted that although \texttt{ASVspoof2021} is often used as a standard dataset to evaluate model generalizability to unseen attacks~\cite{liu2023asvspoof}, part of the real samples in \texttt{ASVspoof2021} originate from the same dataset (the \texttt{VCTK} corpus~\cite{VCTK}) as the \texttt{ASVspoof2019} training data~\cite{todisco2019asvspoof, chintha2020recurrent, tak2021end, xie2023learning, wang2021investigating}. As a result, the real samples from \texttt{ASVspoof2019} and \texttt{ASVspoof2021} share a similar distribution, whereas the \texttt{In-the-wild} and \texttt{MLAAD-EN} samples share nearly no overlap with \texttt{ASVspoof} (further discussion in Appendix~\ref{appendix:3}). Generalization to \texttt{In-the}\texttt{-wild} and \texttt{MLAAD-EN} is therefore more challenging than to \texttt{ASVspoof2021}. The large gains reported by SLIM demonstrates how the style-linguistics mismatch helps with generalization to unseen data.

In Table~\ref{tab:performance_frozen}, we also demonstrate the benefits of introducing Stage 1 
by considering features from SLIM variants as inputs to Stage 2: dependency features, the style and linguistics embeddings ($\text{Enc}_{sty}$ and $\text{Enc}_{ling}$), as well as their combination. 
The architecture of the classification head is kept the same, except for the number of neurons in the input layer. The dependency features outperform the rest on the two out-of-domain datasets, while the subspace embeddings perform better on \texttt{ASVspoof2021}. Simply concatenating the style and linguistics embeddings does not yield significant improvements when compared to benchmark models. 
This suggests that the style-linguistics dependency may not be fully captured by supervised training methods without explicit guidance.

\setlength{\tabcolsep}{1.7pt}
\begin{table}[ht]
\centering
\caption{Detection performance on different deepfake datasets. 
Experiments were repeated three times with different random seeds, and average metric values are reported. \#Param refers to the number of trainable parameters (in millions). For SLIM, we sum up parameters trained at both stages. A few models do not make their code open-source, we therefore include the metrics reported in their papers and skip parameter calculation (N/A). Lowest EERs are bolded per category.}
\label{tab:performance_frozen}
\begin{tabular}{crccccccccc}
\toprule
\multirow{2}{*}{Category}& \multirow{2}{*}{Model} & 
\multicolumn{2}{c}{ASVspoof19} & \multicolumn{2}{c}{ASVspoof21} & \multicolumn{2}{c}{In-the-wild} & \multicolumn{2}{c}{MLAAD-EN} & \multirow{2}{*}{\makecell{\#Param \\ (million)}}\\
\cmidrule(lr){3-4} \cmidrule(lr){5-6} \cmidrule(lr){7-8} \cmidrule{9-10}
& & EER$\downarrow$ & F1$\uparrow$ & EER$\downarrow$ & F1$\uparrow$ & EER$\downarrow$ & F1$\uparrow$ & EER$\downarrow$ & F1$\uparrow$  & \\
\midrule
\multirow{18}{*}{\makecell{\bf{Frozen}\\\bf{frontend}\\ (Section.~\ref{frozen})}} & LCNN~\cite{chintha2020recurrent} & 3.7 & .834 & 25.5 & .197 & 65.6 & .373 & 37.2 & .654 & 4\\
& RawNet2~\cite{tak2021end} & 3.0 & .875 & 22.3 & .213 & 37.8 & .602 & 33.9 & .676 & 4\\
& PS3DT~\cite{yadav2023ps3dt} & 4.5 & $-$ & $-$ & $-$ & 29.7 & $-$ & $-$&$-$ & N/A \\
\cmidrule(lr){2-11}
 & W2V-ASP & 3.3 & .858 & 19.6 & .233 & 30.2 & .705 & 29.1 & .715 & 9 \\
& WLM-ASP & \bf0.3 & .983 & 9.0 & .426 & 25.4 & .751 & 30.3 & .709 & 9 \\
& HUB-ASP & 0.5 & .975 & 15.4 & .289 & 29.9 & .718 & 31.0 & .702 & 9 \\
& W2V-LLGF~\cite{wang2021investigating} & 2.3 & .936 & 9.4 & .402 & 25.1 & .756 & 27.8 & .731 & 10 \\
& W2V-LCNN~\cite{xie2023learning} & 0.6 & $-$ & \bf8.1 & $-$ & 24.5 & $-$ & $-$ & $-$ & N/A \\
 & W2V+WLM & 1.8 & .916 & 22.5 & .203 & 30.3 & .704 & 27.0 & .739 & 9 \\
& W2V+HUB & 0.9 & .956 & 14.2 & .310 & 27.9 & .737 & 27.6 & .732 & 9 \\
& WLM+HUB & 0.8 & .963 & 16.7 & .269 & 29.2 & .724 & 28.5 & .720 & 9 \\
& SSL-Fusion~\cite{yang2024robust} & \bf{0.3} & .981  & 8.9 & .419 & 24.2 & .765 & 26.5 & .739 & 10 \\
\cmidrule(lr){2-11}
& \bf{SLIM variants} & \bf{(ours)} & & & & & & & & \\
& Enc$_{sty}$ & 6.7 & .740 & 8.6 & .438 & 29.2 & .724 & 25.4 & .756 & 9 \\
& Enc$_{ling}$ & 5.9 & .764 & 9.3 & .407 & 30.4 & .708 & 25.0 & .760 & 9 \\
& Enc$_{style+ling}$ & 3.5 & .834 & 9.0 & .429 & 25.1 & .757 & 23.9 & .772 & 10 \\
& Dependency & 2.8 & .897 & 20.5 & .234 & 25.8 & .750 & 19.8 & .811 & 9\\
 & Full & 0.6 & .969 & \bf8.3 & .451 & \bf12.9 & .895 & \bf13.5 & .865 & 11 \\
\midrule
\multirow{3}{*}{\makecell{\bf{Finetuned}\\\bf{frontend}\\(Section.~\ref{ft})}}  & W2V-ASP~\cite{9747768} & 0.3 & .984 & 4.5 & .646 & 18.6 & .836 & 19.2 & .817 & 317 \\
& W2V-AASIST~\cite{tak2022automatic} & \bf0.2 & .991 & \bf{3.6} &.707 & 17.5 & .847 & 14.5 & .856 & 317 \\
& SLIM (ours) & \bf{0.2} & .989 & 4.4 & .651 & \bf12.5 & .898 & \bf10.7 & .892 & 253 \\
\bottomrule
\end{tabular}
\end{table}

\begin{figure}[hbpt]
\centering
\includegraphics[width=0.6\linewidth]{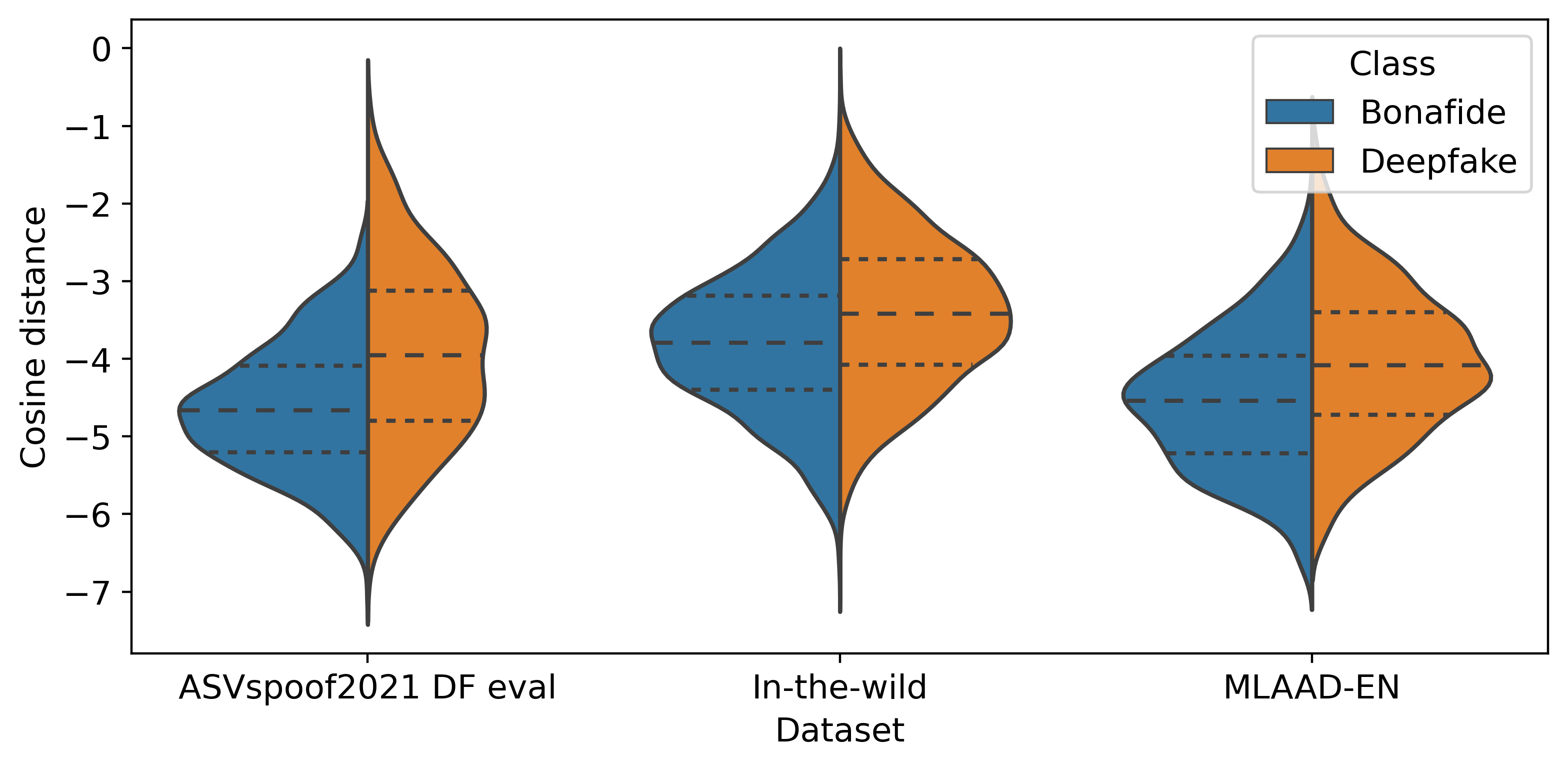}
\caption{Cosine distance (log scale) calculated between the style and linguistics dependency features for \texttt{ASVspoof2021 DF eval}, \texttt{In-the-wild}, and \texttt{MLAAD-EN}. Whiskers from top to bottom represent the 75\% quartile, median, and 25\% quartile of the distribution.}
\label{fig:hist}
\end{figure}

\paragraph{{Style-linguistics mismatch of deepfakes.}} \label{mismatch}
Figure~\ref{fig:hist} shows the distribution of cosine distances between the style and linguistics dependency features for the real and fake classes; larger distances indicate a higher mismatch. Since the distance values approximately follow a Gaussian distribution with unequal variances, we further conduct a Welch's t-test~\cite{ahad2014sensitivity} to examine the statistical significance of the difference between real and fake samples. For all three datasets, the average cosine distance is found to be significantly lower for real speech than for deepfake samples ($p$ < $1\mathrm{e}^{-5}$). This further corroborates our hypothesis that a higher style-linguistics mismatch exists for fakes. On the other hand, the distance distributions of real and fake samples still share a large overlap, indicating that dependency features alone are not sufficient for perfectly discriminating between the two classes. 

\begin{figure}[ht]
    \centering
    \subfigure[]{\includegraphics[width=0.245\textwidth]{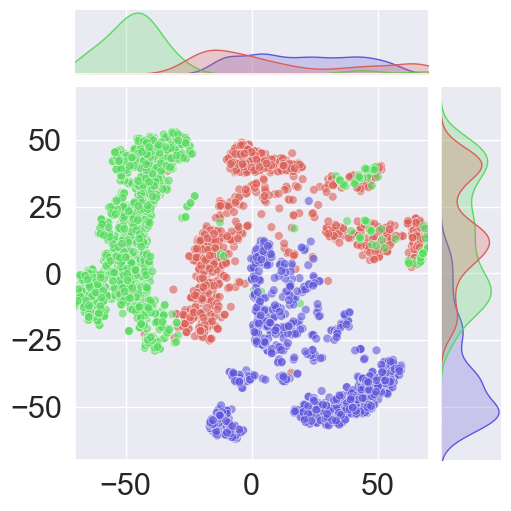}\label{fig:tsne-a}} 
    \subfigure[]{\includegraphics[width=0.245\textwidth]{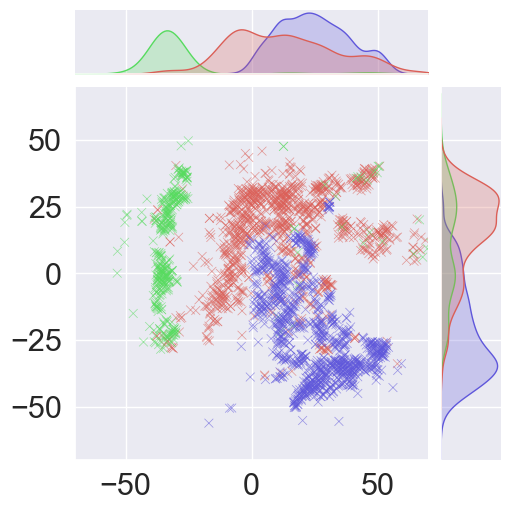}\label{fig:tsne-b}} 
    \subfigure[]{\includegraphics[width=0.245\textwidth]{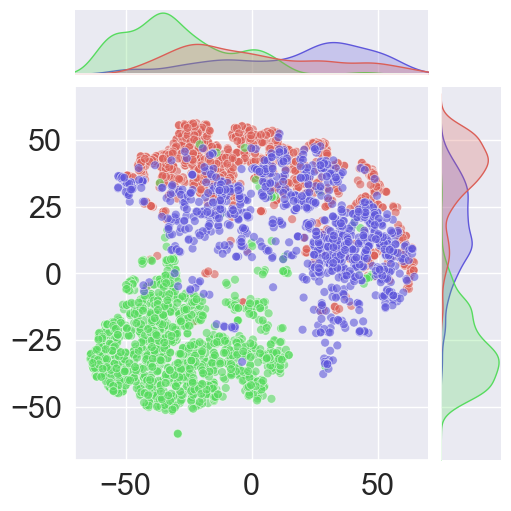}\label{fig:tsne-c}}
    \subfigure[]{\includegraphics[width=0.245\textwidth]{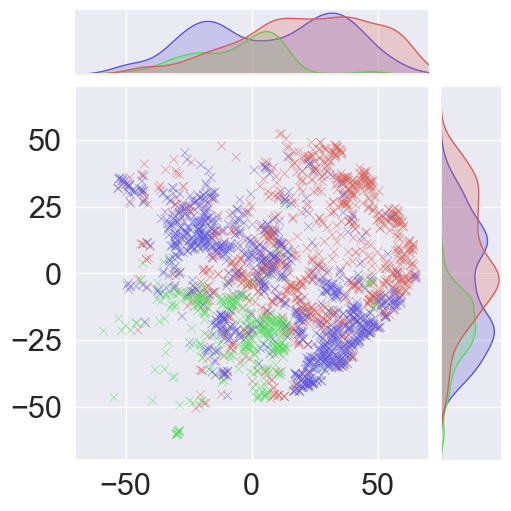}\label{fig:tsne-d}}
    \caption{Projected embeddings using t-SNE for style-linguistic representations: (a) subspace embeddings - real class, (b) subspace embeddings - fake class, (c) dependency features - real class, (d) dependency features - fake class. Data distributions are visualized on the upper and right side of the embedding plots. {\color{red}Red: \texttt{ASVspoof2021}}; {\color{green}Green: \texttt{In-the-wild}}; {\color{blue}Blue: \texttt{MLAAD-EN}}. 
    }
    \label{fig:tsne}
\end{figure}

\paragraph{{Analysis of style-linguistics dependency features.}}
Table~\ref{tab:performance_frozen} demonstrates that style-linguistics dependency features can provide better generalizability than the subspace embeddings (Table~\ref{tab:performance_frozen} SLIM variants, rows 1--4). 
To examine these results, we first aggregate \texttt{ASVspoof2021}, \texttt{In-the-wild}, and \texttt{MLAAD-EN}, and project the dependency features as well as the concatenated subspace embeddings to a 2-dim space using t-SNE for visualization (Figure~\ref{fig:tsne}). Since we use frozen frontends, the embeddings input to Stage 2 training are not affected by backpropagation. Ideal embeddings would exhibit maximal separation between the real and fake classes, while showing minimal shift within each class for different dataset distributions.
In Figure~\ref{fig:tsne}, we see that the dependency features show larger discrimination between real and fakes (\ref{fig:tsne-c} and \ref{fig:tsne-d}) than the concatenated subspace embeddings (\ref{fig:tsne-a} and \ref{fig:tsne-b}), and also a smaller shift between datasets: fake and real samples from the same dataset (color) clusters have less overlap in distribution in the plots.

\begin{figure}[hbpt]
\centering
\includegraphics[width=\linewidth,trim=0 0 0 1.4cm,clip]{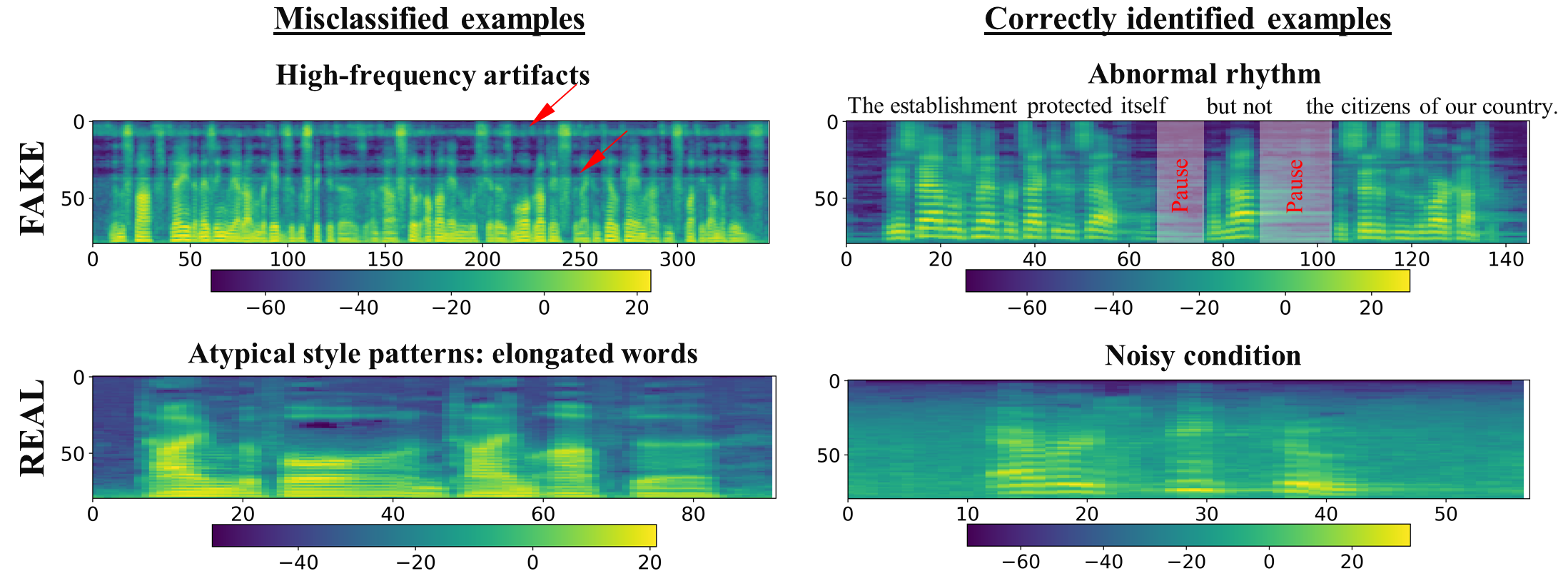}
\caption{
Mel-spectrograms of select samples from \texttt{In-the-wild}. SLIM classifies all four correctly, and when reporting fakes, provides guidance on abnormalities in style and/or linguistics. Also, the dependency and subspace features in SLIM are complementary to each other. Left: samples missed by dependency features but correctly identified by the style and linguistic features; right: vice versa. 
}
\label{fig:example}
\end{figure}

\paragraph{{Interpretation of model decisions.}} \label{explain}
Next, we perform a qualitative evaluation of the model decisions. Figure~\ref{fig:example} shows the mel-spectrograms of four samples selected from \texttt{In-the-wild}.\footnote{These recordings are available in the supplementary documents.} These four demonstrate typical acoustic characteristics that represent a larger group of recordings: (1) top-left is a \emph{fake} sample with audible artifacts at high-frequency region; (2) top-right is a \emph{fake} sample with unnaturally long pauses heard before and after the phrase ``but not''; (3) bottom left is a \emph{real} sample with an atypical speech style where the word pronunciations are elongated; (4) bottom right is a \emph{real} speech recorded in a noisy condition. We find that among the top-performing systems shown in Table~\ref{tab:performance_frozen}, only SLIM classified all four samples correctly (both frozen and fine-tuned versions; with all features), while others mostly failed on (2) and (4). Findings here suggest that SLIM provides guidance when abnormalities in style and linguistics occur. Such guidance can be complemented via \emph{post-hoc} analysis tools such as human evaluations or saliency maps~\cite{arrieta2020explainable} for further interpretation.

Additionally, we note that the decisions made by dependency features and the original subspace representations are complementary to each other. Samples in the right column are correctly identified as fake by the dependency features but missed by the original subspace representations, and vice versa (left column missed by dependency features). These results corroborate with the nature of the two feature types. The dependency features are learned by modelling the general style-linguistics relationship seen in real speech, therefore samples with mismatched style-linguistics pattern are likely to be flagged as ``unreal.'' The original style and linguistics embeddings, on the other hand, are sensitive to signal artifacts, which could be the deepfake imperfections generated during speech synthesis~\cite{shih2024does}, or the amount of background noise and device artifacts. By combining the two features, SLIM captures a variety of abnormalities and achieves improved classification.

\subsection{Ablation studies}
\label{ablation}

\paragraph{\bf{Effects of finetuning SSL frontend.}} \label{ft}
From Table~\ref{tab:performance_frozen}, we see that the frontend finetuning helps to further decrease the EER for SLIM. The finetuned version of SLIM performs better than the rest on (\texttt{In-the-wild},  \texttt{MLAAD-EN}), while providing comparable performance on (\texttt{ASVspoof2019},  \texttt{2021}).
However, it should be noted that the interpretation of style-linguistics mismatch becomes difficult after finetuning, since the two subspace representations may no longer be disentangled.
\paragraph{\bf{Effects of classification backend.}}
In the Stage 2, subspace representations are sent into ASP+MLP layers, which output 256-dim embeddings to fuse with the dependency features. Previous works have shown that different backend architectures may lead to a significant difference in the detection performance~\cite{wang2021investigating}. With the input fixed (dependency features and subspace embeddings), we find that removing the ASP and MLP layers degrades EER across the four datasets (Table~\ref{tab:ab_backend}, Appendix~\ref{appendix:ab}), while using the LCNN~\cite{xie2023learning} or LLGF~\cite{wang2021investigating} backends improves EER on \texttt{ASVspoof2019} and \texttt{ASVspoof2021}, but not on \texttt{In-the-wild} and \texttt{MLAAD-EN}.

\section{Conclusion} \label{conclusion}
We present SLIM, a new ADD framework that models the style-linguistics mismatch to detect deepfakes. Without requiring more labelled data or the added cost of end-to-end finetuning on pretrained encoders, SLIM outperforms existing benchmarks on out-of-domain datasets, while being competitive on in-domain datasets. The learned style-linguistics dependency features are complementary to the individual pretrained style and linguistics subspace representations and also facilitate result interpretation.

\paragraph{Limitations} 
Since our framework explicitly focuses on style-linguistics mismatch, it is possible that real speech samples with atypical style-linguistics dependency (e.g., samples similar Figure~\ref{fig:example} or dysarthric speech~\cite{qian2023survey}) may be misclassified as fakes. One countermeasure is to increase the diversity of real speech in the Stage 1 self-supervised training. Also, although SLIM can benefit from frontend finetuning and more advanced backends, this would affect the feature interpretation and will require modified training approaches. We plan to explore these directions in the future. 

\pagebreak
{
    \small
    \bibliographystyle{plainnat}
    \bibliography{ref}
}

\appendix

\section{Appendix} \label{appendix}

\subsection{Layer-wise analysis of pretrained SSL models} \label{appendix:1}
As mentioned in Section.~\ref{subsec:stage1}, we use the \texttt{Wav2vec-XLSR} model finetuned for emotion recognition (Wav2vec-SER) and speech recognition (Wav2vec-ASR) tasks to extract the style and linguistics representations, respectively. 

\begin{figure}[hbpt]
\centering
\includegraphics[width=0.5\linewidth]{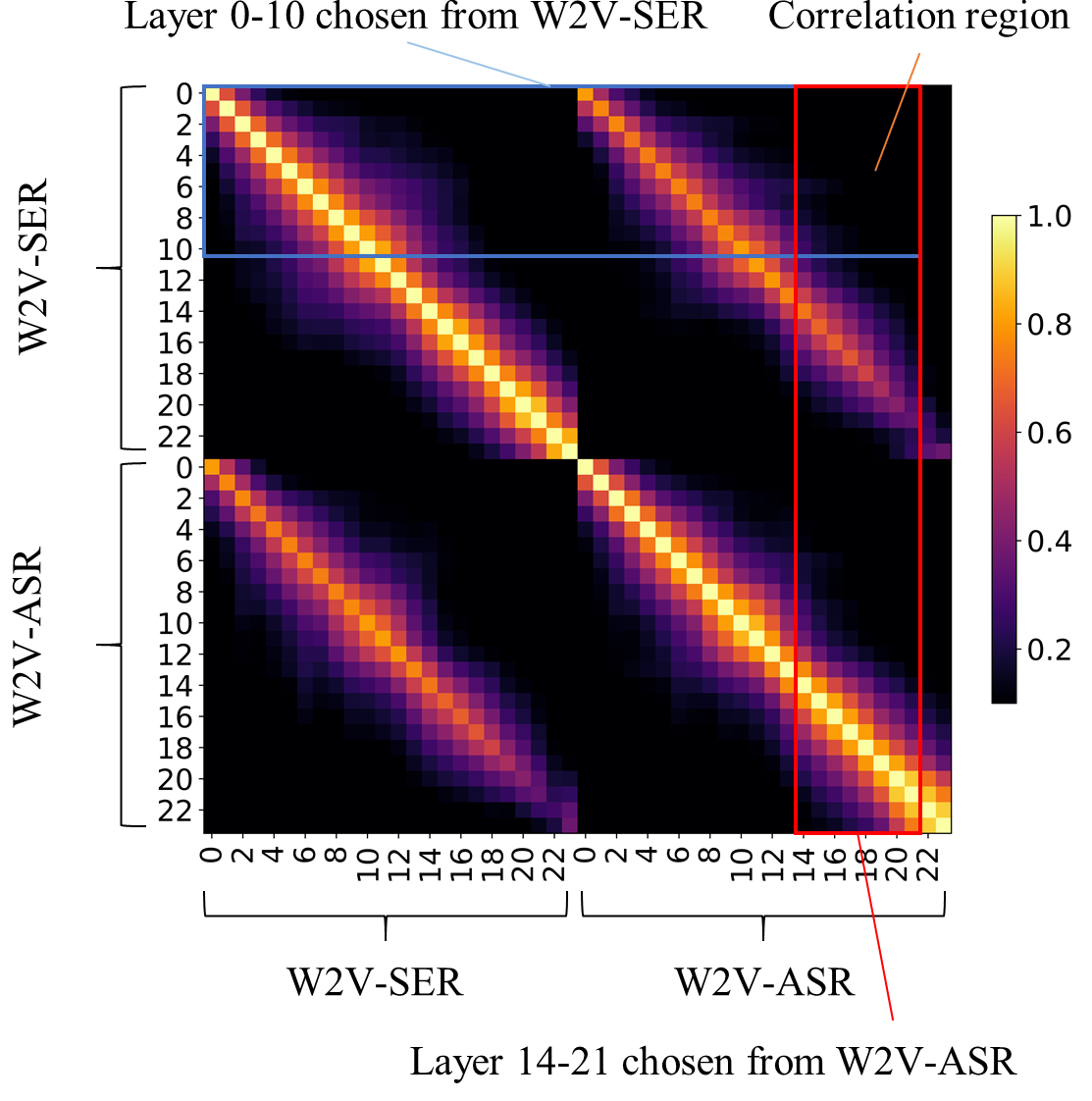}
\caption{Spearman correlation coefficients calculated across all layers from two pretrained Wav2vec-XLSR backbones. {\color{blue}Blue highlights layers 0-10 from Wav2vec-SER to represent style information.} {\color{red}Red highlights layers 14-21 from Wav2vec-ASR to represent linguistics information.} The correlation values between the selected layers can be read from the overlapping region.}
\label{fig:layer}
\end{figure}

The style representation is based on the pretrained model obtained from \url{https://huggingface.co/ehcalabres/wav2vec2-lg-xlsr-en-speech-emotion-recognition} and the linguistics representation is based on the pretrained model obtained from \url{https://huggingface.co/jonatasgrosman/wav2vec2-large-xlsr-53-english}. 
To obtain a maximal disentanglement between the two subspace representations, we calculate Spearman's rank correlation coefficient values between different layers from the two models to examine the layer-wise similarity. These correlation values and our final layer selection are demonstrated in Figure~\ref{fig:layer}. Based on existing works, which showed how linguistics and paralinguistics information propagate through layers, we choose layer 0-10 from Wav2vec-SER backbone to represent style information, and layer 14-21 from Wav2vec-ASR backbone to represent linguistics information. The correlation values between these two groups are shown close to 0, indicating a better disentanglement.

\subsection{Dataset details}
\label{appendix:3}
\begin{figure}[hbpt]
\centering
\includegraphics[width=0.8\linewidth]{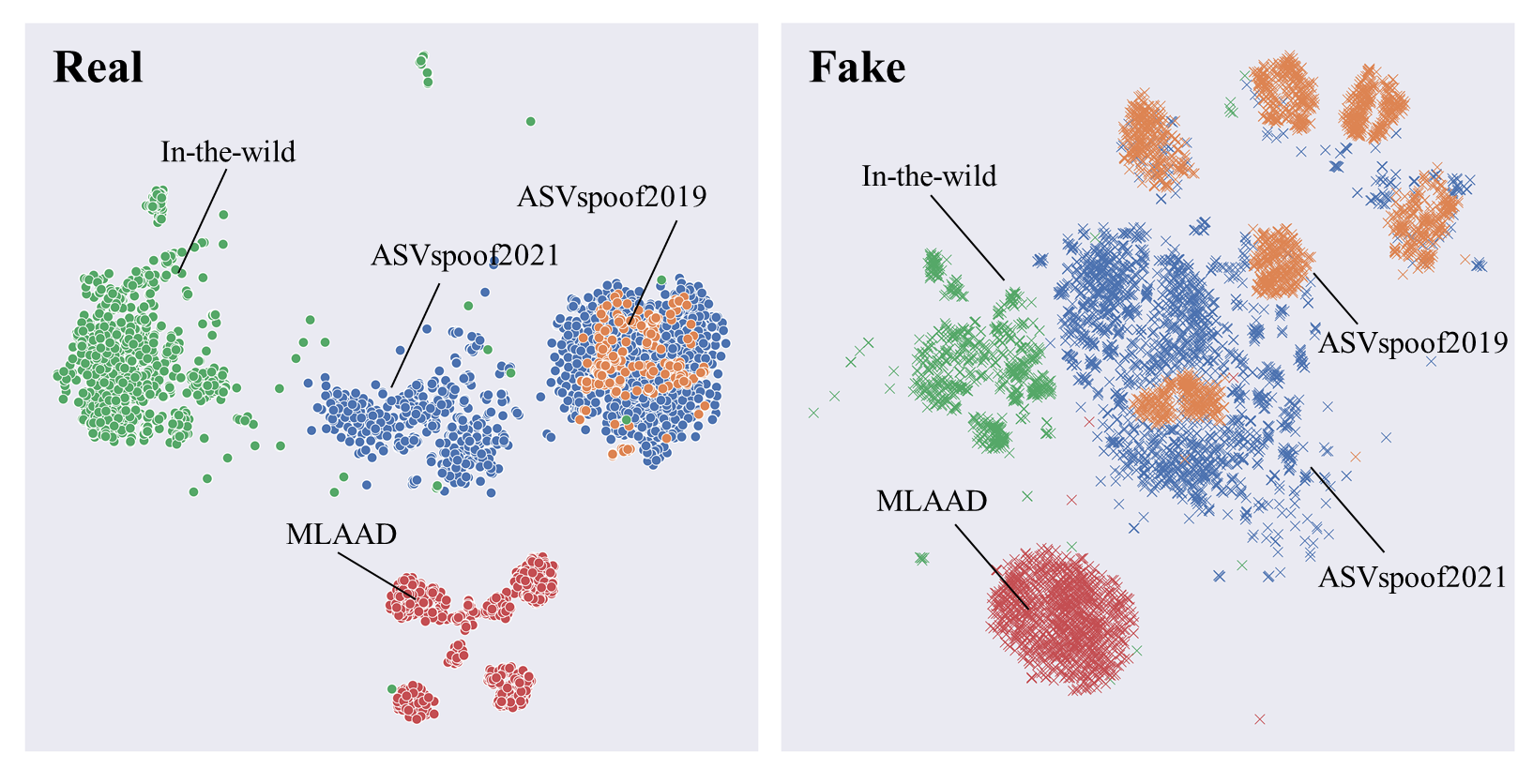}
\caption{Projected WavLM embeddings for real and fake classes from the four employed datasets. Left: real class embeddings. Right: fake class embeddings.}
\label{fig:scatter}
\end{figure}

Table~\ref{tab:dataset} describes the details of datasets used for Stage 1 and Stage 2 training and evaluation. Figure~\ref{fig:scatter} shows the projected WavLM embeddings for real and fake samples from the four employed datasets using t-SNE. We choose WavLM since it is the top-performing model in the single-encoder category (Table~\ref{tab:performance_frozen}). For both classes, an overlap can be seen between \texttt{ASVspoof2019} and \texttt{ASVspoof2021} samples, while samples from \texttt{In-the-wild} and \texttt{MLAAD-EN} can be separated nearly perfectly. This corroborates with the results reported in Table~\ref{tab:performance_frozen} where all employed ADD systems trained on \texttt{ASVspoof2019} perform better on \texttt{ASVspoof2021} than \texttt{In-the-wild} and \texttt{MLAAD-EN}.

\begin{table}[hbpt]
\caption{Summary of datasets used for Stage 1 and Stage 2 training and evaluation.}
\label{tab:dataset}
\centering
\begin{tabular}{lccccccc}
\toprule
\multicolumn{8}{c}{Stage 1 datasets} \\
\midrule 
Name & Split & \#Sample & \#Real & \#Fake & \#Attacks & Speech type & Environment \\
\midrule
\texttt{Common Voice} & Train & 3k & 3k & $-$ & $-$ & Scripted & Crowdsourced \\
\texttt{RAVDESS} & Train & 3k & 3k & $-$ & $-$ & Scripted & Studio \\
\texttt{19 LA train} & Valid & 500 & 500 & $-$ & $-$ & Scripted& Studio \\
\midrule \midrule
\multicolumn{8}{c}{Stage 2 datasets} \\
\midrule
Name & Split & \#Sample & \#Real & \#Fake & \#Attacks & Speech type & Environment \\
\midrule
\texttt{19 LA train} & Train & 25380 & 2580 & 22800 & 6 & Scripted & Studio \\
\texttt{19 LA dev} & Valid & 24884 & 2548 & 22336 & 6 & Scripted & Studio \\
\texttt{19 LA eval} & Test & 71237 & 7355 & 63882 & 17 & Scripted & Studio \\
\texttt{21 DF eval} & Test & 611829 & 22617 & 589212 & 100+ & Scripted & Studio \\
\texttt{In-the-wild} & Test & 31779 & 11816 & 19963 & N/A & Spontaneous & In-the-wild \\
\texttt{MLAAD-EN} & Test & 37998 & 18999 & 18999 & 25 & Scripted & Studio \\
\bottomrule
\end{tabular}
\end{table}

\subsection{Details of the compression module}
\label{appendeix:comp}
\begin{figure}[hbpt]
\centering
\includegraphics[width=0.9\linewidth]{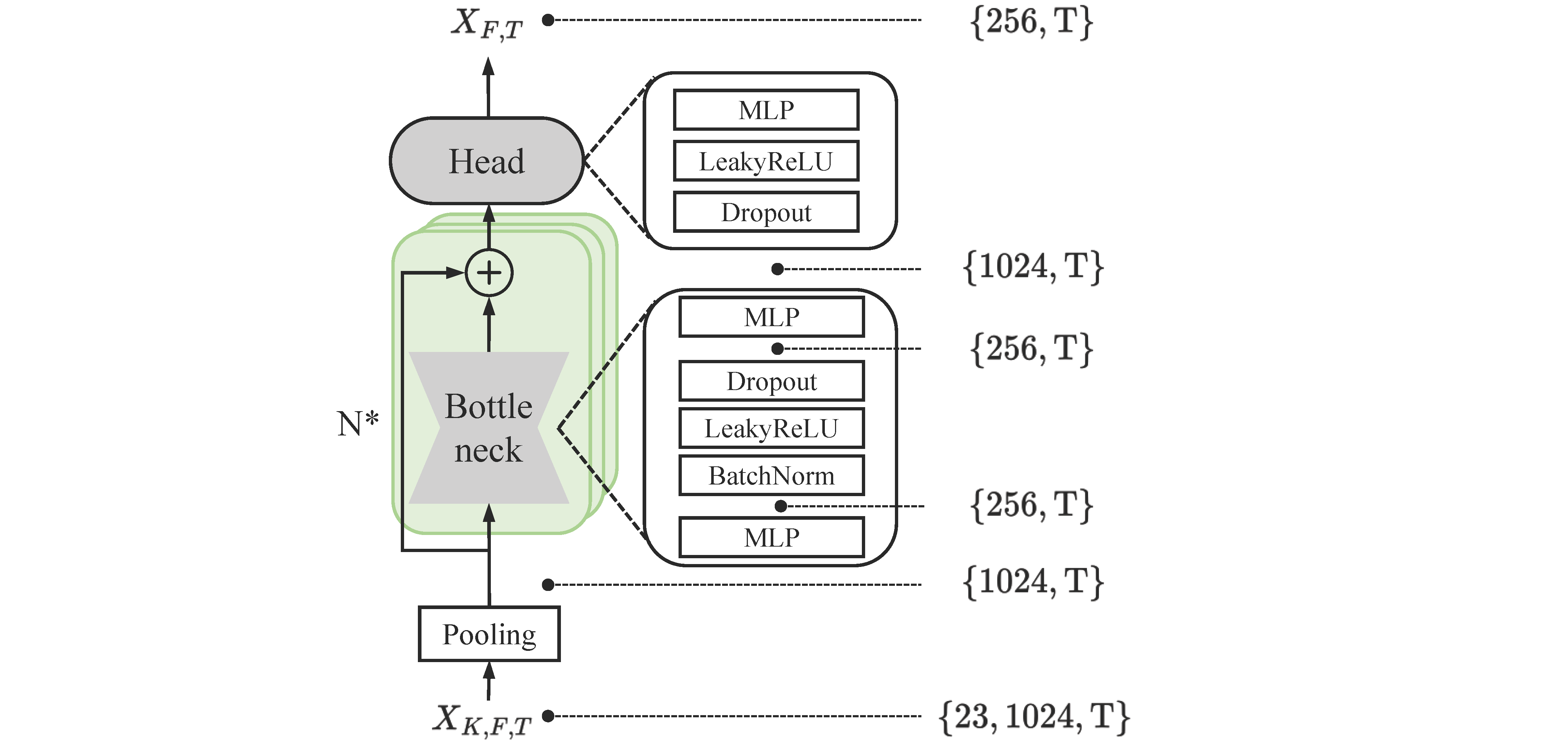}
\caption{Architecture of the compression module with input and output dimensions. Input $\mathbf{X_{K,F,T}}$ represents the original subspace representation encoded by the SSL frontend, where $K$ denotes the transformer layer index, $F$ denotes the feature size, and $T$ denotes the number of time steps.}
\label{fig:compress}
\end{figure}

Figure~\ref{fig:compress} shows the architecture of the compression module. The input is first passed through a pooling layer to obtain an average of different SSL layer outputs. Since the goal of compression modules is to project the original style/linguistics embeddings to a subspace where the compressed embeddings can be maximally correlated, we use bottleneck layers to remove the redundant information that is not shared across the two subspaces. Similar to the design of an autoencoder~\cite{tishby2000information}, the bottleneck layer first compresses the feature dimension from 1024-dim to 256-dim, then recovers it back to 1024-dim. In practice, we found using only one bottleneck layer is enough to obtain meaningful compressed representations. A projection head is applied at the end to reduce the final output dimension to 256.

\subsection{PyTorch implementation of the Stage 1 training objective} \label{appendix:2}
Algorithm~\ref{alg:loss} shows a PyTorch-style implementation of the Stage 1 training objective, which minimizes the cross-subspace distance loss and an intra-subspace redundancy loss. The subspace embeddings are first normalized across the whole batch before passing into the loss calculations. We experimented with two types of distance for the cross-subspace loss: Euclidean and Cosine distance. While no significant difference is found when comparing the performance achieved by the two, the former provides slightly better results, hence is adopted as the final distance measure.

\begin{algorithm}
\SetAlgoLined
\KwIn{$x_{style}$, $x_{linguistic}$}
\KwOut{$L_{stage1}$}
\BlankLine
\tcp{Normalize embeddings from both subspaces}
batch\_size = x\textsubscript{style}.shape[0] \\
x\textsubscript{style\_norm} = torch.nn.BatchNorm1d(x\textsubscript{style}, affine=False) / batch\_size \\
x\textsubscript{linguistic\_norm} = torch.nn.BatchNorm1d(x\textsubscript{linguistic}, affine=False) / batch\_size \\
\BlankLine
\tcp{Computation of cross-subspace loss}
D = torch.linalg.norm(x\textsubscript{style\_norm} - x\textsubscript{linguistic\_norm}, ord='fro') \\
D = torch.pow(D, 2) \\
\BlankLine
\tcp{Computation of intra-subspace redundancy loss}
v\textsubscript{linguistic}  = torch.mm(x\textsubscript{linguistic\_norm}.T, x\textsubscript{linguistic\_norm}) \\
C\textsubscript{linguistic} = torch.linalg.norm(v\textsubscript{linguistic}-torch.eye(v\textsubscript{linguistic}.shape[-1])) \\
C\textsubscript{linguistic}  = torch.pow(C\textsubscript{linguistic}, 2) \\
v\textsubscript{style}  = torch.mm(x\textsubscript{style\_norm}.T, x\textsubscript{style\_norm}) \\
C\textsubscript{style} = torch.linalg.norm(v\textsubscript{style}-torch.eye(v\textsubscript{style}.shape[-1])) \\
C\textsubscript{style}  = torch.pow(C\textsubscript{style}, 2)
\caption{PyTorch-style code for the Stage 1 loss function}
\label{alg:loss}
\BlankLine
\tcp{Final loss term}
L\textsubscript{stage1} = D + $\lambda$ (C\textsubscript{style}+C\textsubscript{linguistic}) \\
\end{algorithm}

\subsection{Performance comparison of different backend classifiers}
\label{appendix:ab}
Table~\ref{tab:ab_backend} shows the performance obtained when the ASP+MLP layers are swapped with other layer choices.

\setlength{\tabcolsep}{4pt}
\begin{table}[ht]
\centering
\caption{Performance comparison of different backend classifiers used in SLIM. Frontend encoders are frozen.}
\label{tab:ab_backend}
\begin{tabular}{ccccc}
\toprule
\multirow{2}{*}{SLIM backend} & \multicolumn{4}{c}{EER} \\
\cmidrule{2-5}
& \texttt{ASVspoof2019} & \texttt{ASVspoof2021} & \texttt{In-the-wild} & \texttt{MLAAD-EN} \\
\midrule
Original (ASP+MLP) & 0.6 & 8.3 & 12.9 & 13.5 \\
None & 0.9 & 9.1 & 13.1 & 13.7 \\
LLGF & 0.4 & 7.5 & 13.5 & 13.0 \\
LCNN & 0.3 & 7.9 & 12.8 & 13.9\\
\bottomrule
\end{tabular}
\end{table}

\subsection{Hyperparameters and computation resources}
\label{appendix:4}
Table~\ref{tab:hparam} describes the optimal hyperparameters and architecture details of SLIM used for Stage 1 and Stage 2 training. The hyperparameter names of the data augmentation modules can be found in SpeechBrain \textit{v1.0.0}.

\begin{table}[hbpt]
\caption{Hyperparameters and architecture details of SLIM.}
\label{tab:hparam}
\centering
\begin{tabular}{cc}
\toprule
{\bf{Parameter}} & {\bf{SLIM}} \\
\midrule \midrule
\multicolumn{2}{c}{\bf{Stage 1 Optimization}} \\
Batch size & 16 \\
Epochs & 50 \\
GPUs & 4 \\
Audio length & 5s \\
Optimizer & AdamW \\
LRscheduler & Linear \\
Starting LR & .005 \\
End LR & .0001 \\
Early-stop patience & 3 epochs \\
$\lambda$ & .007 \\
Training time & 1h\\
\midrule
\multicolumn{2}{c}{\bf{SSL frontend}}\\
Style encoder & Wav2vec-XLSR-SER \\
Style layers & 0-10 \\
Linguistic encoder & Wav2vec-XLSR-ASR \\
Linguistic layers & 14-21 \\
\midrule
\multicolumn{2}{c}{\bf{Compression module}}\\
Bottleneck layers & 1 \\
BN dropout & 0.1 \\
FC dropout & 0.1 \\
Compression output dim & 256 \\
\midrule
\multicolumn{2}{c}{\bf{Stage 2 Optimization}} \\
Batch size & 2 \\
Epochs & 10 \\
GPUs & 4 \\
Audio length & 5s \\
Optimizer & AdamW \\
LRscheduler & Linear \\
Starting LR & .0001 \\
End LR & .00001 \\
Early-stop patience & 3 epochs \\
Training time & 10h \\
\midrule
\multicolumn{2}{c}{\bf{Classifier}} \\
FC dropout & 0.25 \\
\midrule
\multicolumn{2}{c}{\bf{Stage 2 data augmentation}} \\
Num augmentations & 1 \\
Concat with original & True \\
Augment prob & 1 \\
Augment choices & Noise, Reverb, SpecAug \\
SNR\_high & 15dB \\
SNR\_low & 0dB \\
Reverb & RIR noise \\
Drop\_freq\_low & 0 \\
Drop\_Freq\_high & 1 \\
Drop\_freq\_count\_low & 1\\
Drop\_freq\_count\_high & 3 \\
Drop\_freq\_width & .05 \\
Drop\_chunk\_count\_low & 1 \\
Drop\_chunk\_count\_high & 5 \\
Drop\_chunk\_length\_low & 1000\\
Drop\_chunk\_length\_high & 2000\\
\bottomrule
\end{tabular}
\end{table}

\pagebreak

\end{document}